\documentclass{pasj00}

\begin{document}
\SetRunningHead{Doi et al.}{JVN observations of radio-loud NLS1s}
\Received{2006/8/31}
\Accepted{2007/4/9}

\title{Japanese VLBI Network observations\\ of radio-loud narrow-line Seyfert 1 galaxies}

\author{
Akihiro \textsc{Doi},\altaffilmark{1}\thanks{Present Address: The Institute of Space and Astronautical Science, Japan Aerospace Exploration Agency, 3-1-1 Yoshinodai, Sagamihara, Kanagawa 229-8510}\thanks{E-mail: akihiro.doi@vsop.isas.jaxa.jp}
Kenta \textsc{Fujisawa},\altaffilmark{1}
Makoto \textsc{Inoue},\altaffilmark{2}
Kiyoaki \textsc{Wajima},\altaffilmark{3}
Hiroshi \textsc{Nagai},\altaffilmark{4,2}\\ 
Keiichiro \textsc{Harada},\altaffilmark{1} 
Kousuke \textsc{Suematsu},\altaffilmark{1}
Asao \textsc{Habe},\altaffilmark{5}
Mareki \textsc{Honma},\altaffilmark{2,4}\\ 
Noriyuki \textsc{Kawaguchi},\altaffilmark{2,4}
Eiji \textsc{Kawai},\altaffilmark{6}
Hideyuki \textsc{Kobayashi},\altaffilmark{7,2}
Yasuhiro \textsc{Koyama},\altaffilmark{6}\\ 
Hiromitsu \textsc{Kuboki},\altaffilmark{6} 
Yasuhiro \textsc{Murata},\altaffilmark{8,9} 
Toshihiro \textsc{Omodaka},\altaffilmark{10} 
Kazuo \textsc{Sorai},\altaffilmark{5}\\ 
Hiroshi \textsc{Sudou},\altaffilmark{11}
Hiroshi \textsc{Takaba},\altaffilmark{11} 
Kazuhiro \textsc{Takashima},\altaffilmark{12}\\ 
Koji \textsc{Takeda},\altaffilmark{10}
Sayaka \textsc{Tamura},\altaffilmark{9,8}
and
Ken-ichi \textsc{Wakamatsu},\altaffilmark{11}
}
\altaffiltext{1}{Faculty of Science, Yamaguchi University, 1677-1 Yoshida, Yamaguchi, Yamaguchi 753-8512}
\altaffiltext{2}{National Astronomical Observatory of Japan, 2-21-1 Osawa, Mitaka, Tokyo 181-8588}
\altaffiltext{3}{Korea Astronomy and Space Science Institute, 61-1 Whaam-dong, Yuseong, Daejeon 305-348, Korea}
\altaffiltext{4}{Department of Astronomical Science, Graduate University for Advanced Studies,\\ 2-21-1 Osawa, Mitaka, Tokyo 181-8588}
\altaffiltext{5}{Division of Physics, Graduate School of Science, Hokkaido University, N10W8, Sapporo, Hokkaido 060-0810}
\altaffiltext{6}{Kashima Space Research Center, National Institute of Information and Communications Technology,\\ 893-1 Hirai, Kashima, Ibaraki 314-8510}
\altaffiltext{7}{Mizusawa VERA Observatory, 2-12 Hoshigaoka, Mizusawa, Oshu, Iwate 023-0861}
\altaffiltext{8}{The Institute of Space and Astronautical Science, Japan Aerospace Exploration Agency,\\ 3-1-1 Yoshinodai, Sagamihara, Kanagawa 229-8510}
\altaffiltext{9}{Department of Space and Astronautical Science, The Graduate University for Advanced Studies,\\ 3-1-1 Yoshinodai, Sagamihara, Kanagawa 229-8510}
\altaffiltext{10}{Faculty of Science, Kagoshima University, 1-21-30 Korimoto, Kagoshima, Kagoshima 890-0065}
\altaffiltext{11}{Faculty of Engineering, Gifu University, 1-1 Yanagido, Gifu 501-1193}
\altaffiltext{12}{Geographical Survey Institute, 1 Kitasato, Tsukuba, Ibaraki, 305-0811}


\maketitle

\begin{abstract}
We performed phase-reference very long baseline interferometry~(VLBI) observations on five radio-loud narrow-line Seyfert~1 galaxies~(NLS1s) at 8.4~GHz with the Japanese VLBI Network~(JVN).  Each of the five targets (RXS~J08066+7248, RXS~J16290+4007, RXS~J16333+4718, RXS~J16446+2619, and B3~1702+457) in milli-Jansky levels were detected and unresolved in milli-arcsecond resolutions, i.e., with brightness temperatures higher than $10^7$~K.  The nonthermal processes of active galactic nuclei~(AGN) activity, rather than starbursts, are predominantly responsible for the radio emissions from these NLS1s.  Out of the nine known radio-loud NLS1s, including the ones chosen for this study, we found that the four most radio-loud objects exclusively have inverted spectra.  This suggests a possibility that these NLS1s are radio-loud due to Doppler beaming, which can apparently enhance both the radio power and the spectral frequency.
\end{abstract}

\section{Introduction}
Narrow-line Seyfert~1 galaxies (NLS1s), a class of active galactic nuclei~(AGNs), are defined as having the following optical properties: (1)~the full-width at half-maximum~(FWHM) of H$\beta$ is less than 2000~km~s$^{-1}$, (2)~permitted lines are only slightly broader than forbidden lines, and (3)~[OIII]/H$\beta<$3 \citep{Osterbrock&Pogge1985,Pogge2000}.  NLS1s have been extensively studied on both optical and X-ray bands; results show that many of their properties are clearly different to those of the classical Seyfert galaxies.  There is increasing evidence that NLS1s are extreme AGNs, with accretion rates near the Eddington limit (e.g.,~\cite{Pounds_etal.1995,Boroson2002}) onto relatively lower mass ($\sim10^6 \MO$) black holes \citep{Peterson_etal.2000,Grupe&Mathur2004}, although this picture is still under debate.  On the other hand, the radio properties of NLS1s have not been well investigated; to date, there are only two known systematic surveys \citep{Ulvestad_etal.1995,Moran2000}.  From these two surveys, unlike the optical and X-ray studies, the radio data showed little difference between NLS1s and classical Seyfert galaxies.  \citet{Zhou&Wang2002} suggested that there is a scarcity of radio-loud NLS1s, particularly very radio-loud ones (see also \cite{Komossa_etal.2006a}).  Radio loudness, $R$, was conventionally defined as the ratio of 5-GHz radio to {\it B}-band flux densities, with a threshold of $R=10$ separating radio-loud and radio-quiet objects (e.g.,~\cite{Visnovsky_etal.1992,Stocke_etal.1992,Kellermann_etal.1994}).  The reason for the scarcity of radio-loud ($R>10$) NLS1s is still unknown.  The radio-quietness of NLS1s may possibly be related to the suppression of radio jets emanated from accretion disks with high accretion rates \citep{Greene_etal.2006}, as well as X-ray binaries in the {\it high/soft} state (see, e.g.,~\cite{McClintock&Remillard2003} for a review).  

Radio-loud ($R>10$) NLS1s are rare, but they do exist \citep{Siebert_etal.1999,Grupe_etal.2000,Zhou_etal.2003,Whalen_etal.2006,Komossa_etal.2006a,Komossa_etal.2006b}.  One possible idea that could explain the existence of radio-loud NLS1s is that nonthermal jets are associated with NLS1s, and a relativistic effect on these jets influences the radio loudness of NLS1s, as well as the other radio-loud AGN classes.  Hardening of X-ray spectra during rapid X-ray flares of the radio-loud NLS1 PKS~0558$-$504 could arise from the transient spectral dominance of synchrotron emission from relativistically boosted jets \citep{Wang_etal.2001}, similar to the spectra of radio-loud quasars~(e.g.,~\cite{Reeves_etal.1997}).  Observational evidence for the existence of nonthermal jets and Doppler beaming effect on them are required.  If not, the presence of bright radio lobes or a starburst will be needed to explain the radio excess.  

Very long baseline interferometry~(VLBI) is the most powerful tool available for revealing to such properties by direct imaging.  Arcsecond-resolution observations have resolved the structures of only a few NLS1s \citep{Ulvestad_etal.1995,Moran2000}; however, there was insufficient evidence to prove the existence of jets.  VLBI images at milli-arcsecond~(mas) resolutions had been reported for only three radio-quiet NLS1s: MRK~766, AKN~564 \citep{Lal_etal.2004}, and NGC~5506 \citep{Middelberg_etal.2004}; pc-scale radio structures were revealed in these NLS1s.  VLBI observations on a large number of NLS1s, both radio-quiet and radio-loud objects, are crucial if we are to understand the nature of possible highly energetic jet phenomena in these central engines.

We have started VLBI imaging studies on over a dozen NLS1s, including both radio-quiet and radio-loud objects.  It has previously been reported that VLBI observations for the most radio-loud ($R\approx2000$; \cite{Zhou_etal.2003}) object, SDSS~J094857.3+002225, revealed that Doppler-boosted jets are needed to explain observed high brightness temperatures on its radio emissions \citep{Doi_etal.2006a}.  In the present paper, we report our VLBI survey of five radio-loud NLS1s at 8.4~GHz.  In Section~\ref{section:sample}, we outline the reasons for our selection of NLS1s.  In Section~\ref{section:observationanddatareduction}, we describe our observations and data reduction procedures.  In Section~\ref{section:result}, we present the observational results.  In Section~\ref{section:discussion}, we discuss the implications of the results.  In Section~\ref{section:summary}, we summarize the outcomes of our investigation.  Throughout this paper, a flat cosmology is assumed, with $H_0=71$~km~s$^{-1}$~Mpc$^{-1}$, $\Omega_\mathrm{M}=0.27$, and $\Omega_\mathrm{\Lambda}=0.73$ \citep{Spergel_etal.2003}.

\section{Sample}\label{section:sample}
We selected five targets out of the nine radio-loud NLS1s that were previously identified by \citet{Zhou&Wang2002} from 205 NLS1s listed in ``A catalogue of quasars and active nuclei: 10th Ed. \citep{Veron-Citty&Veron2001}.'' The reason for the choice is that we could retrieve the National Radio Astronomy Observatory's~(NRAO's) VLA archival data, obtained at 4.9 or 8.4~GHz with A-array configuration, which provide radio positions with sufficient accuracy for processing in a VLBI correlator.  Our sample of the five radio-loud NLS1s is listed in Table~\ref{table:sample}.

\begin{table}
\caption{Radio-loud NLS1 sample for JVN observations.}\label{table:sample}
\begin{center}
\begin{tabular}{lcccc} \hline\hline
\multicolumn{1}{c}{Name} & $z$ & $S_\mathrm{1.4GHz}^\mathrm{FIRST}$ & $S_\mathrm{5GHz}^\mathrm{VV}$ & $R_\mathrm{5GHz}$ \\
\multicolumn{1}{c}{} &  & (mJy) & (mJy) &  \\
\multicolumn{1}{c}{(1)} & (2) & (3) & (4) & (5) \\\hline
RXS~J08066+7248 & 0.0980  & 49.6\footnotemark[$*$] & 20  & 85  \\
RXS~J16290+4007 & 0.2720  & 11.9  & 22  & 182  \\
RXS~J16333+4718 & 0.1161  & 65.0  & 47  & 105  \\
RXS~J16446+2619 & 0.1443  & 90.8  & 99  & 200  \\
B3~1702+457 & 0.0604  & 118.6  & 26  & 11  \\\hline
\end{tabular}
\end{center}
{\footnotesize Col.~(1) source name; Col.~(2) redshift; Col.~(3) 1.4~GHz flux density from the Faint Images of the Radio Sky at Twenty-centimeters (FIRST; $\sim\timeform{5"}$ resolution; \cite{Becker_etal.1995}); Col.~(4) 5~GHz flux density \citep{Veron-Citty&Veron2001}; Col.~(5) radio loudness in \citet{Zhou&Wang2002}, which were derived from 5~GHz flux density and {\it V}-band magnitude listed in \citet{Veron-Citty&Veron2001} assuming a spectral index of $-0.5$.\\
\footnotemark[$*$] Flux density at 1.4~GHz from the NRAO VLA Sky Survey (NVSS; $\sim\timeform{45"}$ resolution; \cite{Condon_etal.1998}).
}
\end{table}

\section{Observations and data reduction}\label{section:observationanddatareduction}
\subsection{JVN observations}\label{section:JVNobservation}

The five radio-loud NLS1s were observed at 8.4~GHz with the Japanese VLBI Network~(JVN), a newly-established VLBI network, with baselines ranging $\sim50$--2560~km, spread across the Japanese islands (\authorcite{Fujisawa_etal.inprep} in~prep.; \cite{Doi_etal.2006b}).  This array consists of ten antennas, including four radio telescopes of the VLBI Exploration of Radio Astrometry project (VERA; \cite{Kobayashi_etal.2003}).  The observation dates and telescope participants are listed in Table~\ref{table:observationlist}.  Right-circular polarization was received in two frequency bands, 8400--8416~MHz (IF1) and 8432--8448~MHz (IF2), providing a total bandwidth of 32~MHz.  The VSOP-terminal system was used as a digital back-end; digitized data in 2-bit quantization were recorded onto magnetic tapes at a data rate of 128~Mbps.  Correlation processing was performed using the Mitaka FX correlator \citep{Shibata_etal.1998} at the National Astronomical Observatory of Japan.

\begin{table*}
\caption{JVN observations.}\label{table:observationlist}
\begin{center}
\begin{tabular}{llclc} 
\hline\hline
\multicolumn{1}{c}{Date} & \multicolumn{1}{c}{Antenna\footnotemark[$*$]} & $\nu$ & \multicolumn{1}{c}{Target} & $t_\mathrm{scan} \times N_\mathrm{scan}$ \\
\multicolumn{1}{c}{} & \multicolumn{1}{c}{} & (GHz) & \multicolumn{1}{c}{} & (sec) \\
\multicolumn{1}{c}{(1)} & \multicolumn{1}{c}{(2)} & (3) & \multicolumn{1}{c}{(4)} & (5) \\\hline
2006Mar17 & VMI VIR VIS GIF Ks Ud YMG & 8.424 & RXS~J08066+7248 & 130$\times$45 \\
2006Mar26 & VERA$\times$4 GIF Ks TKB YMG & 8.424 & RXS~J08066+7248 & 130$\times$6 \\
 &  &  & RXS~J16290+4007 & 120$\times$22 \\
 &  &  & RXS~J16446+2619 & 120$\times$16 \\
2006May20 & VERA$\times$4 GIF Ks Ud TKB YMG & 8.424 & RXS~J16333+4718 & 158$\times$15 \\
 &  &  & B3~1702+457 & 162$\times$18 \\\hline
\end{tabular}
\end{center}
{\footnotesize Col.~(1) observation date; Col.~(2) antenna participant; Col.~(3) observing frequency at band center; Col.~(4) target name; Col.~(5) scan length in second and number of scans.\\
\footnotemark[$*$] Station code --- Ks: Kashima 34~m of NICT, Ud: Usuda 64~m of JAXA, YMG: Yamaguchi 32~m of NAOJ, TKB: Tsukuba 32~m of GSI, GIF: Gifu 11~m of Gifu University, VMI: VERA Mizusawa 20~m, VIR: VERA Iriki 20~m, VOG: VERA Ogasawara 20~m, and VIS: VERA Ishigaki 20~m of NAOJ.}
\end{table*}

Because the targets in milli-Jansky are too weak for fringe detection with a short integration period, we used a phase-referencing technique that involved fast switching of an antenna's pointing direction.  The switching-cycle period was usually 5 minutes, or $\sim$3 minutes at low elevations.  For three targets~(RXS~J08066+7248, RXS~J16333+4718, and B3~1702+457), we adopted observation schedules for bigradient phase referencing~(BPR; \cite{Doi_etal.2006c}) using two calibrators: $\ldots$-C1-C2-C1-C2-C1-C2-C1-T-C1-T-C1-T-C1-$\ldots$, where C1, C2, and T represent the primary calibrator, the secondary calibrator, and the desired target, respectively.  C1 should be strong enough to be detected in a few minutes.  Even if C2 is a fringe-undetectable calibrator in a few minutes, the BPR can make it a fringe-detectable one by coherent integration of phase-referenced data for several tens of minutes.  The detected C2 will be used as either (1)~an alternative focal point, instead of C1, to reduce the separation angle between a target and the calibrator or (2)~a tracer to measure undesirable phase-drifts in the sky in order to shift the focus to the nearest point from T on the line of C1--C2~\citep{Doi_etal.2006c}.  For B3~1702+457, because the three sources were not in alignment, we used C2 as~(1).  For RXS~J16290+4007, we scheduled two secondary calibrators (as ``C2'' and ``C3'') around the target in order to measure two-dimensional phase-gradients in the sky (cf., \cite{Fomalont&Kogan2005}).  For RXS~J16446+2619, no secondary calibrator was used, because we expected to detect this relatively strong target without BPR.

\begin{table}
\caption{Phase-reference calibrators.}
\begin{center}
\begin{tabular}{llc} 
\hline\hline
\multicolumn{1}{c}{Target} & \multicolumn{1}{c}{Calibrator} & $\Delta \theta$ \\
\multicolumn{1}{c}{} & \multicolumn{1}{c}{} & (deg) \\
\multicolumn{1}{c}{(1)} & \multicolumn{1}{c}{(2)} & (3) \\\hline
RXS~J08066+7248 & J0808+7315\footnotemark[$*$] & 0.46  \\
 & JVAS 0754+7140  & 1.44  \\
RXS~J16290+4007 & J1625+4134\footnotemark[$*$] & 1.56  \\
 & J1623+3909  & 1.50  \\
 & NRAO 512 & 2.23  \\
RXS~J16333+4718 & J1637+4717\footnotemark[$*$] & 0.74  \\
 & J1628+4734 & 0.85  \\
RXS~J16446+2619 & J1642+2523\footnotemark[$*$] & 1.04  \\
B3~1702+457 & J1707+4536\footnotemark[$*$] & 0.67  \\
 & B3 1702+460 & 0.32  \\\hline
\end{tabular}
\end{center}
{\footnotesize Col.~(1)~target's name; Col~(2)~calibrator's name; Col.(3)~separation angle between target and calibrator.\\
\footnotemark[$*$] Primary calibrator as C1 (Section \ref{section:JVNobservation}).}
\end{table}

\subsection{Data reduction}\label{section:datareduction}

Data reduction procedures were performed in following the standard procedure of data inspection, flagging, fringe-fitting, and bandpass calibration using the Astronomical Image Processing System (AIPS; \cite{Greisen2003}) developed at the US National Radio Astronomy Observatory.  A standard a-priori amplitude calibration was not used, mainly because several JVN antennas were not equipped with the monitoring system of system noise temperature, $T_\mathrm{sys}$.  Amplitude-gain parameters relative to each antenna were obtained by self-calibration for a point-like strong source, which was near a target in the sky and scanned every several tens of minutes.  A scaling factor of absolute amplitude was obtained from the result of a-priori calibration using the aperture efficiencies and $T_\mathrm{sys}$ logs of only three antennas~(Yamaguchi 32~m, Kashima 34~m, Usuda 64~m) with the $T_\mathrm{sys}$ monitors.   Such a flux calibration appeared to achieve an accuracy level of 10\% or less, according to several experiments on the JVN.

We obtained correction parameters for both amplitude and phase by self-calibration of C1 with the AIPS task CALIB using a source structure model, which was established in the Difmap software \citep{Shepherd1997} using deconvolution and self-calibration algorithms iteratively.  The correction parameters were applied to the data of T, C2, and C3.  After correcting the positions of the phase-referenced C2 and C3, we derived phase-drift curves from the solutions of self-calibration on both C2 and C3.  The amplitude of the phase-drift curves was appropriately scaled-up/down by factors that should be determined from the ratios of separation angles and position angles of the target-calibrator pairs \citep{Doi_etal.2006c}.  For the observations of RX~J16333+4719, RX~J08066+7248, and RX~J16446+2619, the C2--C1 pair is almost parallel to the T--C1 pair in the sky.  Therefore, we determined the scaling factors so that $\overrightarrow{\rm C_1 C_2^\prime \ } = r_{12} \overrightarrow{\rm C_1 C_2}$, where C2$^\prime$ is the nearest point on the C1--C2 line from T and $r_{12}$ is the scaling factor.  For the data of RXS~J16290+4007 including C2 and C3, we determined two scaling factors so that $\overrightarrow{\rm C_1 T} = r_{12} \overrightarrow{\rm C_1 C_2} + r_{13} \overrightarrow{\rm C_1 C_3}$.  The target and the calibrators in B3~1702+457 observation does not align.  Hence, we applied to T the raw solutions of self-calibration on C2, implying that we obtained a closer reference point by replacing C1 with C2.

Imaging and deconvolution of the calibrated data were carried out using Difmap.  Frequency averaging was done in each IF of 16~MHz, resulting in a field of view of $\sim\timeform{0".3}$ due to bandpass smearing.  We tentatively searched emission components with peak intensities larger than $5\sigma$ of the noise on natural-weighting images in the field of view.  We detected all five targets.  After adjusting the mapping centers to the position of emission peaks, we re-imaged them.  Astrometric measurements were made in these images using the JMFIT of AIPS task.  In addition, we performed self-calibration on four sources only in phase, and obtained solution parameters for all available antennas with signal-to-noise ratios of more than 3.0.  The image dynamic ranges have slightly improved.  Self-calibration could not be performed on RXS~J08066+7248 because of its weakness.  A residual phase-drift in the phase-referenced (i.e., not self-calibrated) data of RXS~J08066+7248 can be estimated from those of the secondary calibrator, JVAS~0754+7140, because this type of phase error is mainly dependent on the separation angle of source pair \citep{Beasley&Conway1995}.  The root-mean-square~(RMS) of phase error in the phase-referenced JVAS~0754+7140 was measured and found to be $\timeform{39D}$; the separation angles of RXS~J08066+7248 and JVAS~0754+7140 from J0808+7315 are $\timeform{0D.46}$ and $\timeform{1D.87}$, respectively.  Therefore, the phase error in RXS~J08066+7248 was estimated to be $\timeform{9D.6}$, causing a coherence loss of only a few percent in amplitude.

\section{Results}\label{section:result}
\begin{figure*}
  \begin{center}
	\includegraphics[width=0.8\linewidth]{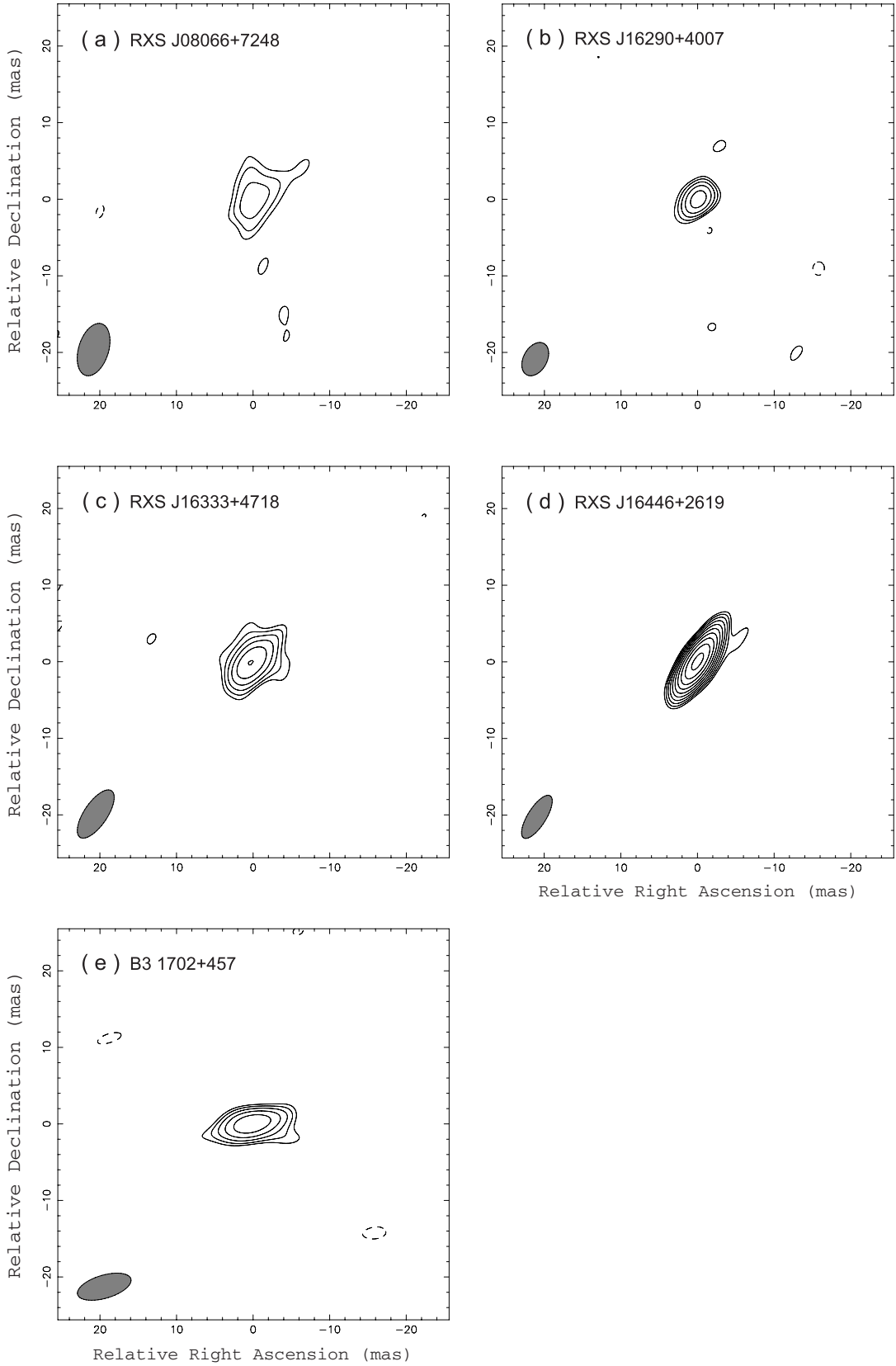}
  \end{center}
  \caption{JVN images of radio-loud NLS1s at 8.4~GHz.  The source name is indicated at upper-left corner in each panel.  Data of all sources, except for RXS~J08066+7248, were self-calibrated~(Section~\ref{section:datareduction}).  All images were synthesized in natural weighting.  Contour levels are separated by factors of $\sqrt{2}$ beginning at 3 times the RMS of image noise~(Table~\ref{table:imageparameter}).  Negative and positive contours are shown as dashed and solid curves, respectively.  Half-power beam sizes (Table~\ref{table:imageparameter}) are given in the lower left corners.}\label{figure:image}
\end{figure*}

We detected each of the five radio-loud NLS1s in mas resolutions, as shown in Fig.~\ref{figure:image}.  These are the first VLBI images for these NLS1s.  A single emission component is seen in each image with dynamic ranges of 7.9--75~(Table~\ref{table:imageparameter}).  Flux measurements were carried out by elliptical-Gaussian fitting to the source profiles using the JMFIT of AIPS task.  Flux densities of 7--150~mJy; the values of radio loudness simply derived from the 8.4~GHz JVN flux densities are still in the radio-loud regime for all the objects, except for B3~1702+457.  Radio powers at the rest frame are listed in Table~\ref{table:result}.

\begin{table*}
\caption{Parameters of JVN images.}\label{table:imageparameter}
\begin{center}
\begin{tabular}{lccccc} \hline\hline
\multicolumn{1}{c}{Name} & $\sigma$ & $\theta_\mathrm{maj}\times\theta_\mathrm{min}$ & $PA$ & $DR$ & $l$ \\
\multicolumn{1}{c}{} & (mJy beam$^{-1}$) & (mas$\times$mas) & (deg) &  & (pc/mas) \\
\multicolumn{1}{c}{(1)} & (2) & (3) & (4) & (5) & (6) \\\hline
RXS~J08066+7248 & 0.54  & $3.9\times7.1$ & $-17$ & 7.7  & 1.8  \\
RXS~J16290+4007 & 1.82  & $3.1\times4.7$ & $-29$ & 15  & 4.1  \\
RXS~J16333+4718 & 0.93  & $3.0\times7.3$ & $-34$ & 16  & 2.1  \\
RXS~J16446+2619 & 1.94  & $2.4\times6.5$ & $-32$ & 75  & 2.5  \\
B3~1702+457 & 0.95  & $3.0\times7.3$ & $-74$ & 16  & 1.2  \\\hline
\end{tabular}
\end{center}
{\footnotesize Col.~(1)~target's name; Col.~(2)~RMS of image noise; Col.~(3)~FWHMs of major and minor axes of synthesized beam; Col.~(4)~position angle of the beam major axis; Col.~(5)~image dynamic range, defined as the ratio of peak intensity to RMS of image noise; Col.~(6)~linear scale in pc corresponding to 1~mas at the distance to the source.}
\end{table*}

\begin{table*}
\caption{Observational results.}\label{table:result}
\begin{center}
\begin{tabular}{lcccccc} \hline\hline
\multicolumn{1}{c}{Name} & \multicolumn{2}{c}{Astrometric position (J2000.0)} & $S_\mathrm{8.4GHz}^\mathrm{VLBI}$ & $I_\mathrm{8.4GHz}^\mathrm{VLBI}$ & $P_\mathrm{8.4GHz}$ & $T_\mathrm{B}$ \\
\multicolumn{1}{c}{} & RA & Dec & (mJy) & (mJy beam$^{-1}$) & (W Hz$^{-1}$) & (K) \\
\multicolumn{1}{c}{(1)} & (2) & (3) & (4) & (5) & (6) & (7) \\\hline
RXS~J08066+7248 & 08 06 38.95744 & 72 48 20.4042 & $6.9\pm1.4$ & $4.2\pm0.7$ & 23.2  & $>10^{7.4}$ \\
RXS~J16290+4007 & 16 29 01.31060 & 40 07 59.9061 & $26.3\pm4.0$ & $27.2\pm3.3$ & 24.6  & $>10^{8.4}$ \\
RXS~J16333+4718 & 16 33 23.58079 & 47 18 58.9298 & $21.2\pm2.9$ & $15.0\pm1.8$ & 23.8  & $>10^{8.0}$ \\
RXS~J16446+2619 & 16 44 42.53399 & 26 19 13.2257 & $150.6\pm15.4$ & $145.5\pm14.7$ & 24.8  & $>10^{9.0}$ \\
B3~1702+457 & 17 03 30.38302 & 45 40 47.1679 & $18.5\pm2.6$ & $15.1\pm1.8$ & 23.2  & $>10^{8.0}$ \\\hline
\end{tabular}
\end{center}
{\footnotesize Col.~(1)~target's name; Col.(2)--(3)~astrometric position, measured relative to C1, by our phase-referenced VLBI observation.  A position uncertainty was 1~mas or less, which was dominated by absolute-position uncertainties of C1 as an ICRF source \citep{Ma_etal.1998,Fey_etal.2004}; Col.~(4)~flux density.  Error was determined as root-sum-square of flux calibration error (10\%; Section~\ref{section:datareduction}) and Gaussian-fitting error (Section~\ref{section:result}); Col.~(5)~peak intensity; Col.~(6) radio power at a rest frequency of 8.4~GHz, in which {\it k}-correction was applied assuming a two-point spectral index derived from VLA at 1.4~GHz~(Table~\ref{table:sample}) and JVN flux density at 8.4~GHz; Col.~(7)~brightness temperature at the rest frame [eq.~(\ref{equation:brightnesstemperature})]. }
\end{table*}

All sources were unresolved in the JVN beams, resulting in brightness temperatures higher than $2.8\times10^7$--$1.1\times10^9$~K at the rest frame~(Table~\ref{table:result}), which were calculated using
\begin{equation}
T_\mathrm{B} = 1.8 \times 10^9 (1+z) \frac{S_\nu}{\nu^2 \phi_\mathrm{maj}\phi_\mathrm{min}} 
\label{equation:brightnesstemperature}
\end{equation}
in K, where $z$ is redshift, $S_\nu$ is the flux density in mJy at frequency $\nu$ in GHz, $\phi_\mathrm{maj}$ and $\phi_\mathrm{min}$ in mas are the fitted full widths at half maximum of the major and minor axes of source size, respectively (cf., \cite{Ulvestad_etal.2005}).  Because these were unresolved, we adopted one-half the beam sizes, i.e., $\theta/2$ (Table~\ref{table:imageparameter}), as the upper limits to the source sizes $\phi$.

The 8.4~GHz JVN flux densities for the two most radio-loud objects in our sample, RXS~J16290+4007 and RXS~J16446+2619, were larger than 1.4~GHz VLA ones~(Tables~\ref{table:sample} and~\ref{table:result}).  Although the difference between the beam sizes of VLA and JVN causes resolution effect, we can obtain at least the lower limit of spectral index $\alpha$ ($S_\nu \propto \nu^{+\alpha}$).  Hence, the two inverted ($\alpha>0$) spectra must be real, without any regard for possible flux variability.

\section{Discussion}\label{section:discussion}
We discuss the origin of the detected radio flux densities and what makes these NLS1s radio-loud, in the present paper.  Possible radio emitting sources with relatively high brightness temperatures in active galaxies are (1)~an accretion disk, (2)~circumnuclear ionized torus, (3)~compact super-nova remnants~(SNRs), and (4)~AGN jets.  

An effective temperature in accretion disk would be at most $10^7$~K even in the innermost region (within several times the Schwarzschild radius) of a ``slim disk,'' a theoretical model for super-Eddington accretion \citep{Abramowicz_etal.1988}, which may be a possible central engine for NLS1s (e.g., \cite{Mineshige_etal.2000,Wang&Netzer2003}).  The emissions from such a small region and a relatively low temperature could hardly account for the detected radio fluxes with JVN.  The nucleus of the classical Seyfert galaxy, NGC~1068, has a component, S1, that may trace thermal free-free emissions from the ionized region of innermost ($\sim$1~pc) molecular torus \citep{Gallimore_etal.2004}.  However, this is an exceptional example and its brightness temperature was at most $\sim10^6$~K, far inferior when compared with the detected radio-loud NLS1s.  

The measured brightness temperatures, $T_\mathrm{B} > 2.8\times10^7$--$1.1\times10^9$~K~(Table~\ref{table:result}), were quite high, which is evidence for the existence of a nonthermal process.  Many VLBI detections have been reported for very young, compact SNRs, for example, SN~1993J \citep{Bietenholz_etal.2001}.  However, even the most luminous radio SNR, SN~1988Z ($z\approx0.022$), generated a radio power at 8.4~GHz of $\sim10^{21.3}$~W~Hz$^{-1}$ at the maximum in its light curve \citep{vanDyk_etal.1993}.  The sources detected with JVN clearly exceed this limit, excluding a compact SNR origin.  Although the radio powers taken from a sample of the most radio-luminous starbursts are $\sim10^{22.3}$--$10^{23.4}$~W~Hz$^{-1}$ \citep{Smith_etal.1998}, these brightness temperatures were derived to be $\lesssim10^5$~K, much less than those of the detected radio-loud NLS1s.  Thus, the stellar origin should be ruled out.    

Previously, based on the sensitivity of $\sim10^6$--$10^8$~K, VLBI imaging has been used to prove that radio emissions associated with AGNs are powered by a nonthermal process related to the activity of a central engine, not only for strong radio AGN classes but also weak radio AGN classes, such as Seyfert galaxies (e.g., \cite{Preuss&Fosbury1983,Neff&Bruyn1983}) and radio-quiet quasars (e.g., \cite{Blundell&Beasley1998,Ulvestad_etal.2005}).  The radio emissions detected from our NLS1 sample are also likely to be related to the activity of central engines.  Nonthermal jets are presumably associated with them, although the JVN images did not resolve any structures.  SDSS~J094857.3+002225, the most radio-loud NLS1, has been resolved with VLBI into multiple radio components with very high brightness temperatures requiring Doppler boosting \citep{Doi_etal.2006a}, indicating highly relativistic jets.  Possible jet structures have also been found with VLBI in NGC~5506 \citep{Middelberg_etal.2004}, a radio-quiet NLS1 candidate \citep{Nagar_etal.2002}.  In the limited dynamic ranges, the JVN images have presumably shown an unresolved core as the base of jets or one of the compact hot-spots in radio lobes in these radio-loud NLS1s.

\begin{figure}
  \begin{center}
	\includegraphics[width=\linewidth]{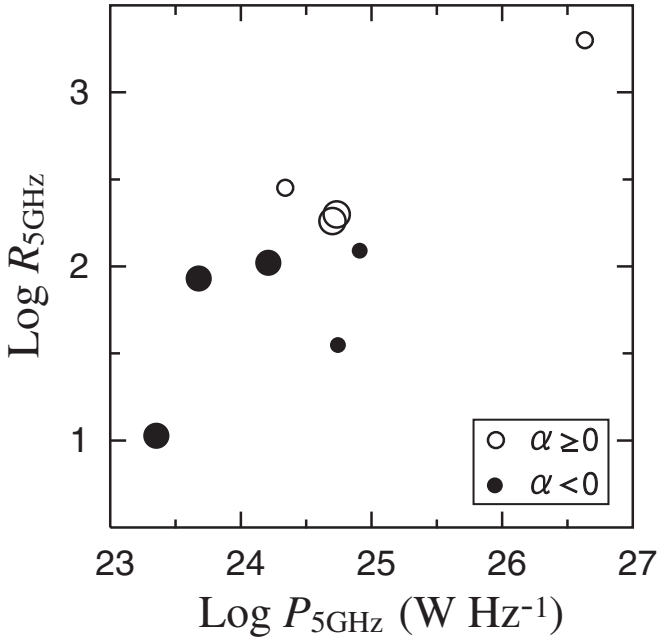}
  \end{center}
  \caption{Radio loudness vs.~5 GHz radio power for the sample of radio-loud NLS1s from \citet{Zhou&Wang2002} and \citet{Doi_etal.2006a}.  The radio data were taken mainly from \citet{Veron-Citty&Veron2001}, see Section~\ref{section:result} in detail.  Filled and open circles represent objects with steep spectrum and inverted spectrum, respectively ($S_\nu \propto \nu^{+\alpha}$).  Five large plots represent objects observed with JVN in the present study.}\label{figure:RL-P5GHz}
\end{figure}

We now discuss the relationship between the radio loudness and the radio spectral index.  We made the plot of radio loudness vs.~radio power using (a)~the nine radio-loud NLS1s listed in \citet{Zhou&Wang2002}, including the five NLS1s observed with JVN (see, Section~\ref{section:sample}), and (b)~the most radio-loud narrow-line quasar SDSS~J094857.3+002225 ($R\approx 2000$; \cite{Zhou_etal.2003}), as shown in Fig.~\ref{figure:RL-P5GHz}.  RXS~J00449+1921, one of the sample in \citet{Zhou&Wang2002}, was excluded from the plot because recent observations have found it to be radio-quiet \citep{Maccarone_etal.2005}.  Total flux densities at 1.4--5~GHz were taken from \citet{Veron-Citty&Veron2001} for these NLS1s, except for HE~0132$-$4313 at 4.85--8.4~GHz \citep{Grupe_etal.2000}, PKS~0558$-$504 at 2.7--4.85~GHz \citep{Wright&Otrupcek1990,Wright_etal.1994}, and SDSS~J094857.3+002225 at 1.43--4.86~GHz \citep{Doi_etal.2006a}.  We discovered that the four inverted spectrum sources are the four most radio-loud objects: SDSS~J094857.3+002225 with $R\approx 2000$, 2E~1640+5345 with $R=282$, RXS~J16446+2619 with $R=200$, and RXS~J16290+4007 with $R=182$.  This suggests that there could be some connection between a strong low-frequency absorption and the high values of radio loudness.  We discuss two possibilities to cause the combination of the high radio loudness and the strong absorption.  (1)~One possibility is Doppler beaming effect on jets.  The radio flux density from jets could have been boosted by a factor of $\delta^{3-\alpha}$ (Doppler factor), where $\delta \equiv \sqrt{1-\beta^2}/(1-\beta\cos{\phi})$, $\beta \equiv v/c$ ($v$ is the source speed), and $\phi$ is the angle between the direction of the source velocity and our line of sight, although the optical emission from accretion disk is not affected: higher radio loudness could be achieved.  The peak frequency of a self-absorbed synchrotron spectrum could be enhanced by a factor of $\delta$.  The frequency range where an inverted spectrum could be seen would extend to our observing frequency.  Doppler beaming in an NLS1 has already been established by variability \citep{Zhou_etal.2003} and VLBI studies \citep{Doi_etal.2006a} for SDSS~J094857.3+002225, the most radio-loud object in Fig.~\ref{figure:RL-P5GHz}.  (2)~Another possibility is a very compact radio lobe.  Giga-hertz peaked spectrum objects~(GPSs; \cite{ODea1998} for a review) are strong, compact radio sources, and thought to be in a very early stage~($<10^3$~yr) on the evolution of radio galaxies.  The radio lobes of GPSs are probably self-absorbed due to high brightness, and its spectral evolution throughout the evolution of radio galaxy has been suggested (e.g., \cite{Snellen_etal.2000}).  According to the evolution framework, radio lobes become more luminous and less absorbed with age.  However, although we used the limited number of radio-loud NLS1s, we cannot find any evidence of such an expected tendency in Fig.~\ref{figure:RL-P5GHz}: inverted (i.e., strongly absorbed) spectra are rather seen at a high radio power regime.  It is less likely that radio-loud NLS1s are a kind of GPSs.  Therefore, we suggest the possibility that Doppler boosting has affected the radio loudness of these NLS1s showing inverted spectra.  

We also have radio-loud NLS1s with steep ($\alpha<0$) spectra.  At least our VLBI detections have revealed the existence of components with high brightness temperatures in the three steep spectrum radio-loud NLS1s, as well as the two inverted spectrum ones.  In case of $\alpha=-0.6$, more than about half of the 1.4~GHz VLA flux densities (Table~\ref{table:sample}) have been retrieved with JVN at 8.4~GHz, implying that a compact nonthermal component made a major contribution toward total radio fluxes.  However, we have little suggestion about the reason why they are radio-loud.  There may be following possibilities.  (1)~Doppler beaming may exist but mildly affected: there would be an insufficient boosting on frequency, but a sufficient boosting on radio flux to the observables, due to only $\delta$-times boosting on frequency, but $\delta^{3-\alpha}$-times boosting on flux density.  (2)~These NLS1s may have significant jet structures that can provide a radio power sufficient to being radio-loud but cannot be resolved at the spatial resolution of JVN, $\sim3$~mas$\times7$~mas.  In the condition of equipartition between synchrotron electrons and magnetic fields, a diameter of about 0.4~mas or more would be needed for a component size in these NLS1s so that a jet can be optically-thin (i.e., steep spectrum) at frequencies higher than 1.4~GHz.  Even if radio lobes with a size of 1~mas or less possibly resided in these NLS1s, they could not be resolved with JVN beams.  We cannot ruled out either of the possibilities only by the present study.  

We have carried out another VLBI observation for the same sample at 1.7~GHz with the US Very Long Baseline Array~(VLBA), and the results will be reported in a future paper.  We expect that radio properties of the radio-loud NLS1s will be revealed in more detail, because optically-thin, extended synchrotron emissions would be detected more easily at such a low frequency.

\section{Summary}\label{section:summary}
We observed five radio-loud NLS1s at 8.4~GHz with the Japanese VLBI Network~(JVN) using a phase-referencing technique.  All the targets were detected and unresolved in mas resolutions, i.e., with brightness temperatures higher than 10$^7$--10$^9$~K.  VLBI-detected flux densities kept four out of the five sources still radio-loud.  Radio powers mainly originate in the nonthermal processes of AGN activity in central engines, rather than starbursts.  We argued the properties of nonthermal jets in these NLS1s.  The two most radio-loud objects in our sample, RXS~J16290+4007 and RXS~J16446+2619, showed inverted spectra between VLA flux densities at 1.4~GHz and JVN ones at 8.4~GHz.  With nine radio-loud NLS1s, we also found that the four most radio-loud objects exclusively have inverted spectra.  We suggest there is a possibility that the radio emissions of these NLS1s are enhanced by Doppler beaming, which can change both radio loudness and the peak frequency of synchrotron self-absorption spectra.

\bigskip

The JVN project is led by the National Astronomical Observatory of Japan~(NAOJ) that is a branch of the National Institutes of Natural Sciences~(NINS), Hokkaido University, Gifu University, Yamaguchi University, and Kagoshima University, in cooperation with the Geographical Survey Institute~(GSI), the Japan Aerospace Exploration Agency~(JAXA), and the National Institute of Information and Communications Technology~(NICT).  We have made use of NASA's Astrophysics Data System Abstract Service, the NASA/IPAC Extragalactic Database (NED), which is operated by the Jet Propulsion Laboratory; it also has made use of Ned Wright's on-line cosmology calculator.  The National Radio Astronomy Observatory is a facility of the National Science Foundation operated under cooperative agreement by Associated Universities, Inc.


\end{document}